\documentclass[%
reprint,
superscriptaddress,
amsmath,amssymb,
prl,
]{revtex4-2}

\usepackage{graphicx}% Include figure files
\usepackage{dcolumn}% Align table columns on decimal point
\usepackage{bm}% bold math

\renewcommand{\vec}[1]{\ensuremath{\boldsymbol{#1}}}

\begin{document}

\preprint{APS/123-QED}

\title{Onset of sliding across scales:\\ How the contact topography impacts 
frictional strength}% Force line breaks with \\

\author{Fabian Barras}
\email{fabian.barras@alumni.epfl.ch}
\affiliation{Civil Engineering Institute, Institute of Materials 
Science and Engineering, Ecole Polytechnique F\'{e}d\'{e}rale de Lausanne 
(EPFL), Station 18, 1015 Lausanne, Switzerland}
\affiliation{The Njord Centre, Department of Physics, Department of Geosciences, University of Oslo, 0316 Oslo, Norway}
%\altaffiliation[Also at ]{Physics Department, XYZ University.}
\author{Ramin Aghababaei}
\affiliation{Engineering Department, Aarhus University, 8000 Aarhus C, Denmark}
\author{Jean-Fran\c{c}ois Molinari}
\affiliation{Civil Engineering Institute, Institute of Materials 
Science and Engineering, Ecole Polytechnique F\'{e}d\'{e}rale de Lausanne 
(EPFL), Station 18, 1015 Lausanne, Switzerland}
\date{\today}% It is always \today, today,

\begin{abstract}
When two solids start rubbing together, frictional sliding initiates in the wake of slip fronts propagating along their 
surfaces in contact. This macroscopic rupture dynamics can be successfully mapped on the elastodynamics of a moving shear 
crack. However, this analogy breaks down during the nucleation process, which develops at the scale of surface asperities where microcontacts form. Recent atomistic simulations revealed how a characteristic junction size selects if the failure of microcontact junctions either arises by brittle fracture or by ductile yielding. This work aims at bridging these two complementary descriptions of the onset of frictional slip existing 
at different scales. We first present how the microcontacts failure observed in 
atomistic simulations can be conveniently ``coarse-grained'' using an equivalent 
cohesive law. Taking advantage of a scalable parallel implementation of the 
cohesive element method, we study how the different failure mechanisms of the microcontact asperities
interplay with the nucleation and propagation of macroscopic slip fronts along the interface. Notably, large 
simulations reveal how the failure mechanism prevailing in the rupture of the 
microcontacts (brittle versus ductile) significantly impacts the nucleation of 
frictional sliding and, thereby, the interface frictional strength. This work 
paves the way for a unified description of frictional interfaces connecting the 
recent advances independently made at the micro- and macroscopic scales.
\end{abstract}

%Use showkeys class option if keyword%display desired
\maketitle

%\tableofcontents

%\include{onset_of_slip}

\section{Introduction}

The rapid onset of sliding along frictional interfaces is often driven by a 
similar dynamics than the one observed during the rupture of brittle materials. 
Just like a propagating shear crack, slipping starts and the shear stress 
drops in the wake of a \textit{slip front} that is moving along the interface. This 
analogy particularly suits the observed behaviors of frictional interfaces 
at a macroscopic scale and explains that the earthquake dynamics has been 
studied for decades as the propagation of shear cracks along crustal 
faults \cite{richards_dynamic_1976, 
kostrov_principles_1988,scholz_mechanics_2010,rosakis_intersonic_2002}. 

Recent experiments \cite{svetlizky_classical_2014} quantitatively demonstrated 
how Linear Elastic Fracture Mechanics (LEFM) perfectly describes the evolution 
of strains measured at a short distance from the interface during the dynamic
propagation of slip fronts. From this mapping, a unique parameter emerges, 
the equivalent fracture energy $G_c$ of the frictional interface, which was 
later used to rationalize the observed arrest of slip fronts in light of the 
fracture energy balance criterion \cite{kammer_linear_2015,bayart_fracture_2016}. The 
same framework was also successfully applied to describe the failure of 
interfaces after coating the surface with lubricant 
\cite{svetlizky_brittle_2017}. Despite a reduction in the force required to 
initiate sliding, the equivalent fracture energy measured after lubrication was 
surprisingly higher than for the dry configuration \cite{bayart_slippery_2016}. 
This apparent paradox in the framework of LEFM is expected to arise during the 
nucleation phase, which is controlled by the microscopic nature of friction and 
contact. At the microscale, surfaces are rough and contact only occurs between 
the surface peaks, resulting in a very heterogeneous distribution of the sliding 
resistance \cite{dieterich_direct_1994,yastrebov_infinitesimal_2015}.

A class of laboratory-derived friction models \cite{dieterich_modeling_1979,ruina_slip_1983,marone_laboratory-derived_1998} 
has been successfully used to rationalize some key aspects of the rupture nucleation along frictional interfaces, particularly in the context of earthquakes (critical length scales at the onset of frictional instabilities \cite{ruina_slip_1983,rice_stability_1983,rice_rate_2001,ampuero_earthquake_2008,aldam_critical_2017}, speed and type of the subsequent ruptures 
\cite{zheng_conditions_1998,gabriel_transition_2012,brener_unstable_2018-1,barras_emergence_2019,barras_emergence_2020}). The so-called \textit{rate-and-state} formulations are empirically calibrated to reproduce the subtle evolution of friction observed during experiments \cite{dieterich_direct_1994}. A direct connection with the physics of the microcontacts and their impact on the frictional strength remains however unsettled and motivates the recent effort to derive physics-based interpretations of the rate-and-state friction laws 
\cite{estrin_model_1996,baumberger_physical_1999, bar-sinai_velocity-strengthening_2014,aharonov_physics-based_2018}. 

To rationalize the friction coefficient of metal interfaces, 
Bowden and Tabor \cite{bowden_area_1939,bowden_friction_2001} suggested that the 
microcontact junctions represent highly confined regions yielding under a 
combination of compressive and shear stresses. Later, Byerlee 
\cite{byerlee_theory_1967} proposed an alternative for brittle materials, by 
assuming that slipping does not occur through the plastic shearing of junctions 
but rather by fracturing the microcontacts, which leads to a smaller value of 
the friction coefficient in agreement with the ones measured for rock 
interfaces. From atomistic calculations, Aghababaei \textit{et al.} 
\cite{aghababaei_critical_2016,aghababaei_debris-level_2017,aghababaei_origins_2019} recently derived a 
characteristic size of the microcontact junction $d^*$ controlling the transition 
from brittle fracture (of junctions larger than $d^*$) to ductile yielding (of junctions smaller than $d^*$). As sketched 
in Fig.~\ref{fig:setup}, these brittle and ductile failure mechanisms co-exist along two rough surfaces 
rubbing together. From this permanent interplay, Fr\'{e}rot \textit{et al.} 
\cite{frerot_mechanistic_2018} proposed a new interpretation of surface wear 
during frictional sliding, while Milanese \textit{et al.} 
\cite{milanese_emergence_2019} discussed the origin of the self-affinity of 
surfaces found in natural or manufactured materials. The link between these 
different microcontact failure mechanisms and the macroscopic frictional 
strength of the interface remains however overlooked.

In this work,  we first present how to approximate the microcontacts failure using a convenient cohesive model. 
The cohesive approach is then implemented in a high-performance finite element 
library and used to simulate the onset of sliding across two scales. At the 
macroscopic level, we study the ability of an interface to withstand a 
progressively applied shearing, i.e. its frictional strength, while at the 
microscopic scale, we observe how the failure process develops across the 
microcontact junctions. This study culminates by discussing how small differences 
in the interface conditions or the size of asperity junctions, only visible at 
the scale of the microcontacts, can nevertheless have a significant impact on 
the nucleation phase and the macroscopic frictional strength. 

%\subsection{}
%\subsubsection{}
\section{Problem description \label{sect:onset:geometry}}
\begin{figure}
\centering
\includegraphics[width=\linewidth]{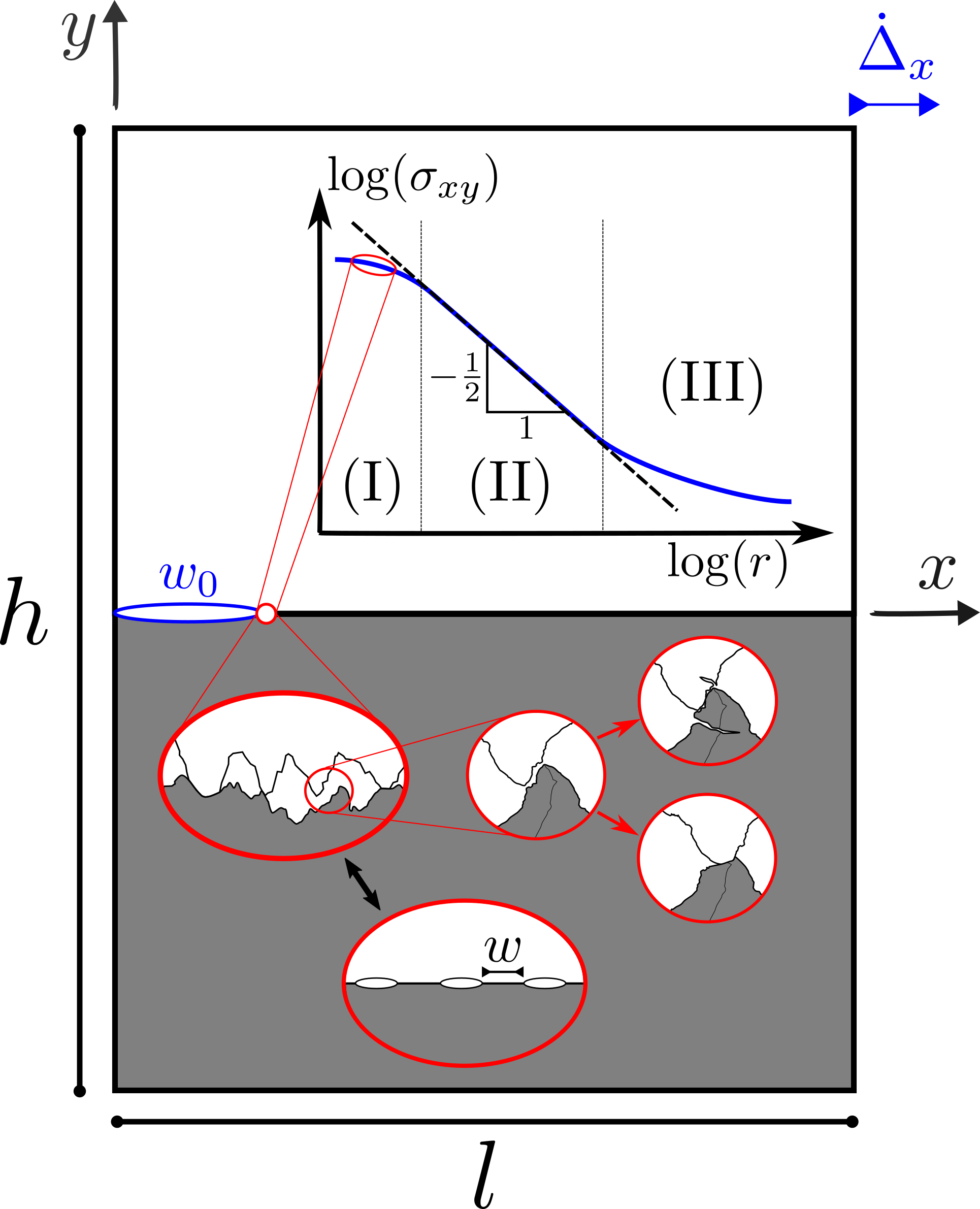}
\caption{\label{fig:setup} Geometry of the problem. The inset presents 
the schematic shear stress $\sigma_{xy}$ profile predicted by LEFM at a 
distance $r$ from a macroscopic rupture front. A nonlinear region 
(I) exists at the immediate vicinity of the tip, followed by a linearly 
elastic region (II), where $\sigma_{xy}$ is dominated by the square root 
singularity. Further away from the tip (III), non-singular contributions 
dominate the profile of $\sigma_{xy}$, which converges toward the far-field 
stress conditions. At the onset of sliding, the microcontacts within the 
nonlinear region (I) can either break by brittle fracture of their apexes 
or by plastic yielding 
\cite{bowden_area_1939,byerlee_theory_1967,aghababaei_critical_2016}. Our work 
aims at describing how these different failure mechanisms occurring at the 
scale of asperity contact, i.e. ``hidden'' within (I), impact the onset of 
sliding and the frictional strength.}
\end{figure}
We consider two linearly elastic blocks of height $h/2$ brought into contact 
along their longitudinal face of length $l$. As presented in Fig.~\ref{fig:setup}, the two blocks are progressively sheared 
by displacing the top surface at a constant speed ${\dot{\Delta}}_x$, while the bottom surface is 
clamped. In a Cartesian system of coordinates, whose origin stands at the left 
edge of the contacting plane, the boundary conditions of this elastodynamic 
problem correspond to 
\begin{equation}
\left\lbrace
\begin{aligned}
&\vec{u}(x,-h/2,t) = 0\\
&{\dot{u}}_x(x,h/2,t) = {\dot{\Delta}}_x\\
&u_y(0,y,t)= u_y(l,y,t)=0
\end{aligned}
\right.
\label{equ:BC}
\end{equation}
and lead to a state of simple shear, for which the shear components of 
the Cauchy stress tensor are $\sigma_{xy}=\sigma_{yx}=\tau$. In Eq.~(\ref{equ:BC}), $\vec{u}=\{u_x,u_y\}$ corresponds to the 
displacements vector and $\dot{\square}$ denotes a time derivative. The elastodynamic solution of this system in the absence of interfacial slip is presented in Fig.~\ref{fig:step_loading} of the Appendix. As illustrated in Fig.~\ref{fig:setup}, sliding nucleates at small scales from the rupture of the microcontacts  which potentially stems from several non-linear phenomena (cleavage, plasticity, interlocking).  As discussed by Aghababaei \textit{et al.} \cite{aghababaei_critical_2016}, atomistic models are particularly suited to simulate these phenomena in comparison to continuum approaches, but are conversely disconnected from the macroscopic dynamics.  Therefore, we rely on a 2D plane strain continuum description of the two solids, while the complex interface phenomena and associated dissipative processes are assumed to be constrained at the contact plane and entirely described by a ``coarse-grained'' cohesive law deriving from a thermodynamic potential $\Phi$. The shape of $\Phi$ and its associated exponential cohesive law correspond to a generic failure response of the microcontact asperities observed during a large set of atomistic simulations \cite{aghababaei_critical_2016,aghababaei_debris-level_2017,aghababaei_asperity-level_2018,aghababaei_origins_2019,milanese_emergence_2019,aghababaei_effect_2019,brink_adhesive_2019}. As sketched in Fig.~\ref{fig:setup}, sliding is assumed to initiate at the edge of a critical nucleus (e.g. the largest non-contacting region or the result of underlying stochastic processes \cite{brener_unstable_2018-1,de_geus_how_2019}) existing at the very left of our model interface with a size $w_0$. Moreover, the rough contact topography sketched in Fig.~\ref{fig:setup} is idealized as a regular pattern of contacting and non-contacting junctions of microscopic size $w\ll w_0$.

Additional details about theoretical derivations, the numerical method and the material properties used in this manuscript are provided in Appendix, which namely defines the values of the Young's modulus $E$, the Poisson's ratio $\nu$ and a reference interface fracture energy $G_c^{\rm ref}$.

\section{Characteristic length scales of the brittle-to-ductile failure transition}

\begin{figure*}
\includegraphics[width=\linewidth]{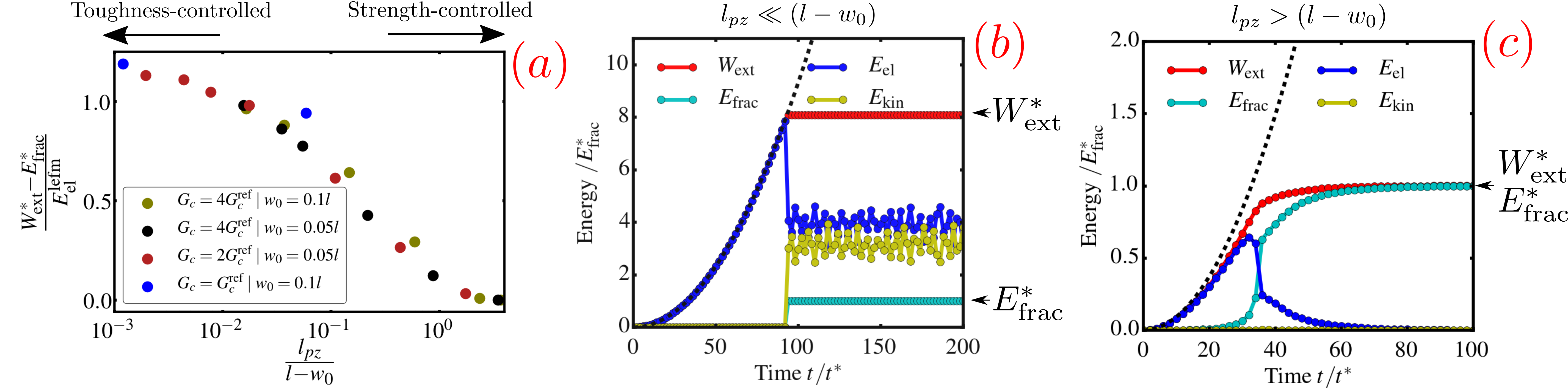}
\centering
\caption{\label{fig:homogeneous} The ratio of the process zone size to the 
length of the junction mediates the work required to initiate 
sliding. (a) Normalized external work required to initiate sliding along a single uniform junction as function of the ratio between the process zone size $l_{pz}$ and the resisting junction size $(l-w_0)$ for different types of interface 
properties and geometries. (b) and (c) present the evolution of energies during the onset of sliding, which occurs respectively at 
$t=92t^*$ and $t=35t^*$. The two events share the same elastic properties 
and $G_c=4 G_c^{\rm ref}$, but their respective interface cohesive laws 
lead to $l_{pz}/(l-w_0)=3.5\cdot10^{-2}$ and $l_{pz}/(l-w_0)=3.5$. The dashed lines in (b) and (c) present 
the build-up of elastic strain energy in the absence of interfacial slip discussed in the Appendix.}
\end{figure*}

Next, we study the onset of slip along a uniform and 
homogeneous interface (i.e. a unique junction) of fracture energy $G_c$ and size $(l-w_0)$. Figure~\ref{fig:homogeneous}b presents the 
evolution of energies observed during a typical failure event, i.e, the applied 
external work $W_{\textmd{ext}}$, the elastic strain energy $E_{\textmd{el}}$, 
the energy dissipated by fracture $E_{\textmd{frac}}$, and the kinetic energy 
$E_{\textmd{kin}}$. During an initial phase, the elastic strain energy builds up 
in the system following the dynamics predicted in the absence of interfacial slip (Fig.~\ref{fig:step_loading}) and depicted by the black dashed line. After an initial loading phase, sliding nucleates at $x=w_0$, a propagating slip front breaks the 
interface cohesion and releases $E^*_{\textmd{frac}}=G_c(l-w_0)$. The asterisk marks in Figs.~\ref{fig:homogeneous}b-c simply distinguish the final value of energy obtained after the complete interface failure from its transient value, i.e. 
$E_i^*=E_i(t\gg t^*)$. After the complete failure, an eventual excess of 
mechanical energy ($W^*_{\textmd{ext}}-E^*_{\textmd{frac}}$) remains in the 
system and takes the form of elastic vibrations in absence of any other 
dissipative process. 

Figure~\ref{fig:homogeneous}c describes the evolution of energies 
observed during another failure event, during which sliding initiates for a 
significantly lower applied external work, exactly balancing the energy dissipated in fracture 
($W^*_{\textmd{ext}}=E^*_{\textmd{frac}}$). Perhaps surprisingly to some readers, these quantitatively different sliding events arise within two systems having identical elastic properties ($E$, $\nu$) and interface fracture energy $G_c$. These different dynamics emerge 
solely from the size of the fracture process zone at the tip of the crack which can be estimated as \cite{palmer_growth_1973,onate_analytical_2008}:
\begin{equation}
l_{pz} \cong e\frac{\delta_c}{\tau_c}\frac{E}{(1-\nu^2)} = 
\frac{G_c}{\tau_c^2}\frac{2\mu}{1-\nu}.
\label{equ:ldz}
\end{equation}
$\tau_c$ and $\delta_c$ are respectively the maximum shear strength and 
critical slip displacement entering the cohesive formulation 
(see Eqs.~(\ref{equ:exponential_law}) and (\ref{equ:fracture_energy})). When the size of the process zone $l_{pz}$ is comparable to the
junction size $(l-w_0)$, the sliding motion develops along a damage band 
stretching over the entire length of the interface with an energy balance similar 
to the one observed in Fig.~\ref{fig:homogeneous}c. Conversely, if 
$l_{pz}\ll(l-w_0)$, sliding initiates in the form of a slip front propagating 
from $x=w_0$ and leading to a more violent rupture as described in Fig.~\ref{fig:homogeneous}b. 
The two different stress profiles existing prior to the rupture events presented in Fig.~\ref{fig:homogeneous}b and c can be visualized in Fig.~\ref{fig:lpz} of the Appendix. In the limit of an infinitesimally small process zone, the rupture corresponds to a singular shear (mode II) crack, whose propagation initiates according to LEFM energy balance. In this context, the applied external work should not solely balances $E^*_{\rm frac}$ but also load the system above the strain energy required to initiate the rupture. The latter is derived in the Appendix and can be estimated as ($\chi\approx 1.12$):
\begin{equation}
E^{\textmd{lefm}}_{\textmd{el}} = 
\frac{G_c}{\chi^2}\frac{hl}{\pi w_0(1-\nu)}.
\label{equ:lefmpot}
\end{equation}

For different interface properties and dimensions, Fig.~\ref{fig:homogeneous}a presents how the process zone size (Eq.~(\ref{equ:ldz})) together with the rupture energy balance can rationalize the observed transition from the dynamics of sharp crack-like events (for 
$l_{pz}\ll(l-w_0)$) to gradual ductile failures (for $l_{pz}\geqslant(l-w_0)$). 

In some applications, the system is preferably described in terms of the macroscopic force $F^*_{\textmd{ext}}$ required to trigger sliding, i.e. to reach the interface frictional strength. As presented in Fig.~\ref{fig:force}, the brittle-to-ductile transition can be similarly characterized from the evolution of the force required to initiate sliding between $F^{\rm lefm}_{\textmd{ext}}$ and $F^{\rm str}_{\textmd{ext}}$. Using Eqs (2) and (\ref{equ:F_lefm}), $F^{\rm str}_{\textmd{ext}}$ can be rewritten as

\begin{equation}
F^{\rm str}_{\textmd{ext}} = F^{\rm lefm}_{\textmd{ext}}\Big(\frac{l_{pz}}{l-w_0}\Big)^{-\frac{1}{2}}\sqrt{\chi^2\pi\frac{w_0}{l}\Big(1-\frac{w_0}{l}\Big)}.
\label{equ:F_str2}
\end{equation}

This expression is depicted by the black and gray solid lines in Fig.~\ref{fig:force} and predicts well the evolution of the frictional force observed when the process zone is large. With very small process zones, the frictional force saturates at the value predicted by brittle fracture theory in Eq.~(\ref{equ:F_lefm}).

The evolution between these two failure mechanisms reported in Figs.~2a and \ref{fig:force} is analogous to the transition discussed in the tensile failure of concrete structures \cite{bazant_scaling_1997} from the plastic failure of small specimens to the brittle failure of larger structures. Two important differences arise during the shear failure of frictional interfaces. Brittle and ductile mechanisms co-exist during the failure of rough surfaces and the characteristic length scale is not purely a bulk property but also depends on interface conditions (for example lubrication). Indeed, an equivalent brittle-to-ductile transition exists in the failure of the microcontact asperities observed in the atomistic simulations. Aghababaei \textit{et al.} \cite{aghababaei_critical_2016} revealed how a 
characteristic junction size
\begin{equation}
d^* = \lambda \frac{G_c}{\tau_c^2}\mu
\label{equ:onset:dstar}
\end{equation}
mediates this transition from the brittle rupture of the apexes of junctions larger than $d^*$ to the ductile yielding of 
junctions smaller than $d^*$. In Eq.~(\ref{equ:onset:dstar}), $\lambda$ is a dimensionless factor accounting for the 
geometry (typically in the range of unity) and, therefore, $l_{pz}$ (Eq.~(\ref{equ:ldz})) corresponds to the 
same characteristic length scale than $d^*$ (Eq.~(\ref{equ:onset:dstar})). Remarkably, there is a direct analogy between the 
brittle-to-ductile failure transition (controlled by $d^*$) observed during the failure of microcontact 
asperities \cite{aghababaei_critical_2016} and the failure of the ``coarse-grained'' junctions (controlled by $l_{pz}$) 
presented in Fig.~\ref{fig:homogeneous}a using the cohesive approach. The latter represents therefore a powerful tool to 
unravel the impact of the microcontacts failure on the macroscopic frictional strength of multi-asperity interfaces.

Next, we select two types of interface properties with the same fracture energy $G_c = G_c^{\rm ref}$ and with process zone sizes that are much smaller than the size of the domain. We later refer to these two systems as \textit{interface A} ($l_{pz,A}/l=9\cdot10^{-4}$) and \textit{interface B} ($l_{pz,B}/l=4.5\cdot10^{-2}$). For the single-junction interfaces considered in this section, the interfaces \textit{A} and \textit{B} rupture with a crack-like dynamics (as $l_{pz}\ll l-w_0$) at similar magnitudes of external work (see the blue circles in Fig~\ref{fig:homogeneous}a, which are recalled in Fig~\ref{fig:onset:multicontacts}a). In the next section, the frictional strength of multi-asperity interfaces is studied in light of the characteristic junction size $d^*$. The size of the microcontact junctions $w$ is chosen in order to discuss the cases where $w$ is respectively larger/smaller than the characteristic junction size of the interfaces \textit{A}/\textit{B} ($d^*_A<w<d^*_B$). The characteristic junction sizes are computed using $\lambda\cong3$ in Eq.~(\ref{equ:onset:dstar}), such that $d^*\equiv l_{pz}$. This value of $\lambda$ corresponds to the one estimated for three-dimensional spherical asperities in \cite{aghababaei_critical_2016}.

\section{Rough contact topography and frictional strength}

\begin{figure*}[t!]
\includegraphics[width=\linewidth]{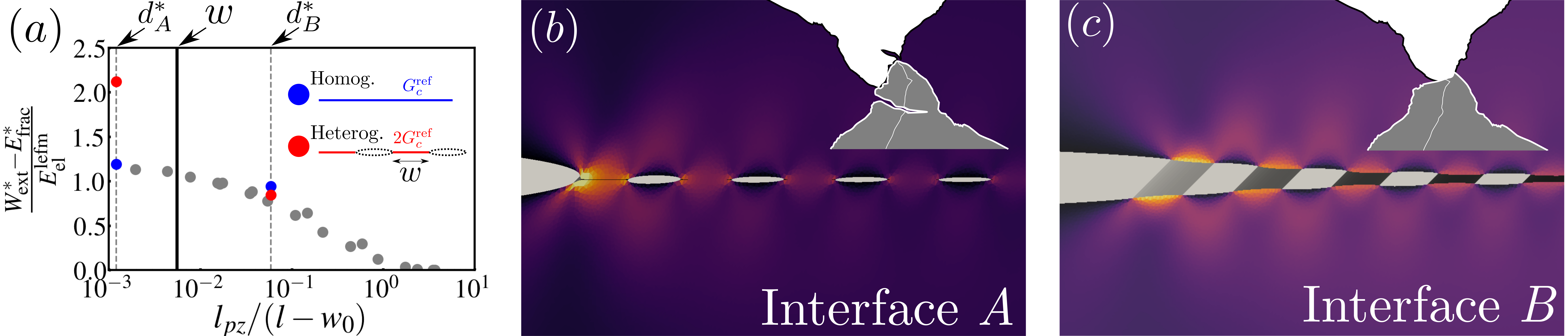}
\centering
\caption{\label{fig:onset:multicontacts} Evolution of the frictional strength in the presence of microcontact junctions for two representative interfaces differentiated by their respective characteristic junction size ($d^*_A<w<d^*_B$). (a) The grey circles recall the data discussed previously in Fig.~\ref{fig:homogeneous}a. The blue circles corresponds to homogeneous (single junction) interfaces. The red circles are associated to multi-asperity interfaces with a heterogeneous microstructure but the same average fracture energy $G_c^{\rm ref}$. (b)-(c) Zooms at the vicinity of the critical nucleus $(x=w_0)$ revealing the origin of the frictional strength difference between interface \textit{A} and \textit{B} in the presence of microcontacts. Colors depict the shear stress profile existing before the onset of sliding while an artificial vertical displacement $\Big(u_y(x,y)=u_x(x,y)\Big)$ is applied to help visualizing the slip profile along the interface ($200$ times magnification). The evolution of junctions strength is depicted with a gradation from black ($\tau^{\mathrm{str}}=\tau_c$) to white ($\tau^{\mathrm{str}}=0$). The sketches located in the top right of each plot associate the failure of the coarse-grained multicontacts interfaces \textit{A} and \textit{B} to the corresponding failure mechanism of surface asperities discussed in Fig.~\ref{fig:setup}.}
\end{figure*}

As sketched in Fig.~\ref{fig:setup}, two solids come into contact along a reduced portion of the interface, between the peaks of the microscopically rough surfaces. To model the effect of this heterogeneous topography, we now introduce an idealized array of microscopic gaps and junctions of size $w=0.05w_0=0.005l$. In order to keep the total energy dissipated into fracture unchanged ($E^*_{\textmd{frac}}=G_c^{\rm 
ref}(l-w_0)$), the fracture energy of the microscopic junctions is set to $2G_c^{\rm ref}$. The interfaces \textit{A} and \textit{B} have significantly different frictional strength in presence of the heterogeneous microstructure as shown by the red circles in Fig.~\ref{fig:onset:multicontacts}a for the external work and in Fig.~\ref{fig:force} for the external force. This major difference is caused by the introduction of a new length scale $w$ in the systems, which exactly stands between the characteristic length scales $d^*_{A}$ and $d^*_{B}$. 

As presented in Fig.~\ref{fig:onset:multicontacts}c, along interface \textit{B} ($d^*_{B}>w$), several microcontact junctions start damaging and slipping during the initial loading phase. The stress concentration at the edge of the critical nucleus spans several microcontact junctions and gaps. Their individual properties are thereby homogenized within this large 
process zone and result in a quasi-homogeneous frictional response driven by the strength-dominated ductile failure. Conversely, for interface \textit{A} ($d^*_{A} < w$), the shear stress sharply concentrates at the very edge of the microcontact junctions (cf. Fig.~\ref{fig:onset:multicontacts}b) whose local toughness directly controls the onset of failure. 

For interface \textit{B}, the effective fracture energy corresponds to the average value which explains that the heterogeneous and homogenized interfaces break at the same magnitudes of $W^{*}_{\textmd{ext}}$ and $F^{*}_{\textmd{ext}}$. For interface \textit{A}, the toughness of the microcontact junctions ($G_c=2G^{\rm ref}_c$) directly controls the failure. From Eqs.~(\ref{equ:Elefm}) and (\ref{equ:F_lefm}), the external work and the external force are hence expected to increase by respectively a factor $2$ and $\sqrt{2}$, in good agreement with the simulated values (reported in Figs.~\ref{fig:onset:multicontacts} and \ref{fig:force}). Such toughening mechanism can therefore become stronger if a larger contrast exists between the toughness of individual microcontacts and the average macroscopic toughness of the interface.

\section{Subsequent rupture dynamics}

\begin{figure*}
\centering
\includegraphics[width=\linewidth]{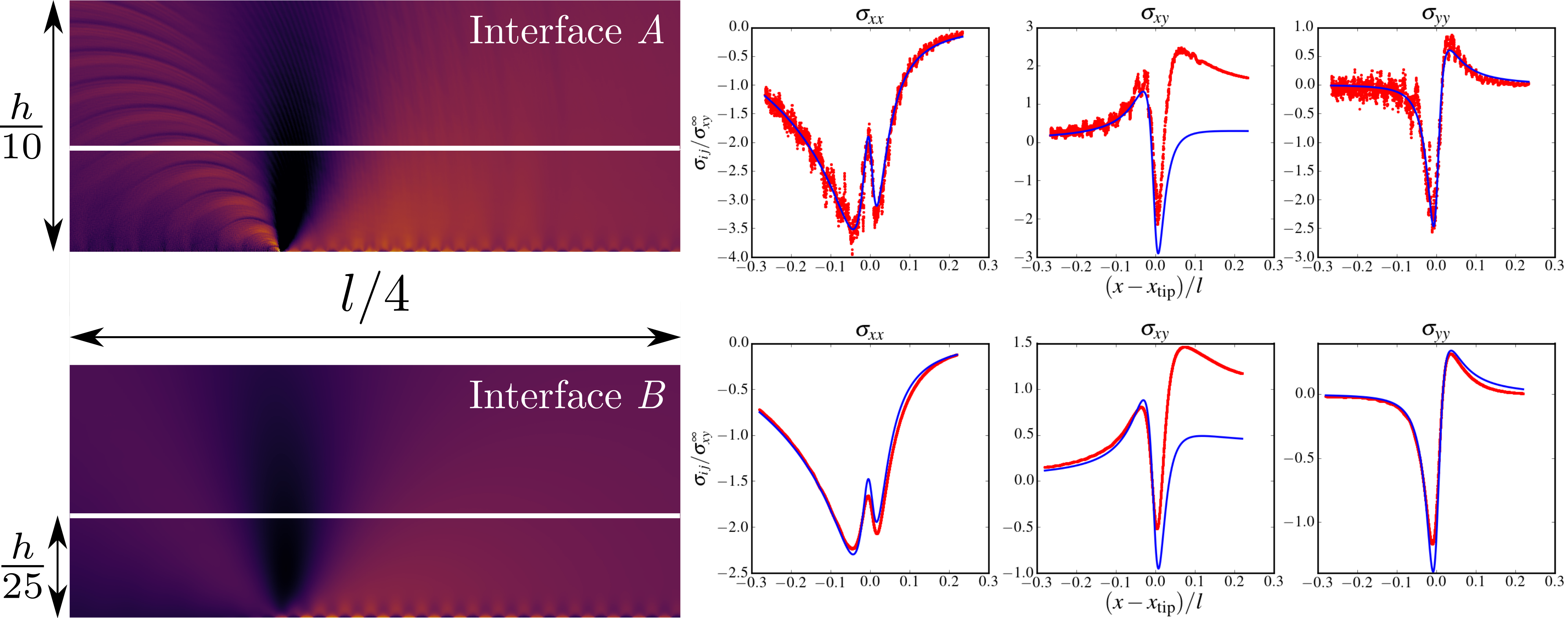}
\caption{\label{fig:heterogeneous} At a macroscopic distance from the interface the 
evolutions of the stress fields observed during the dynamic failure of the heterogeneous interfaces \textit{A} 
(top) and \textit{B} (bottom) comply with LEFM predictions for an interface fracture energy 
corresponding to the average value $G_c^{\rm ref}$. On the left panels, 
shear stress at the vicinity of the propagating slip front is mapped using the 
same color scale. To mimic the experimental measurements, the white lines 
highlight the position along which the components of the Cauchy stress tensor 
are presented on the right panels in red. The stress fields predicted by LEFM at 
the vicinity of a shear crack are plotted in blue for a fracture energy equal to 
$G_c^{\rm ref}$. Note that the mismatch visible in the simulation profiles 
of $\sigma_{xy}$ is caused by the shear wave traveling ahead of an accelerating shear crack which is not included in LEFM 
solutions of Eq.~(\ref{equ:modeII_stress_sing}) \cite{andrews_rupture_1976-1,svetlizky_properties_2016}.}
\end{figure*}

The main objective of the manuscript is to study the impact of the microscopic roughness at nucleation. It is nevertheless insightful to briefly comment the subsequent rupture dynamics observed along the heterogeneous interfaces \textit{A} and \textit{B}. As shown in the previous sections, the details of the microstructure plays an important role during the nucleation phase as the macroscopic frictional strength cannot be systematically predicted from the average interface properties. However, the subsequent rupture dynamics are macroscopically similar and comply with LEFM predictions for homogenized interface 
properties. In Fig.~\ref{fig:heterogeneous}, the stress profiles are measured at a 
macroscopic distance ($h/25\gg w$) from the contacting plane as it is the case 
during experiments \cite{svetlizky_classical_2014,bayart_fracture_2016,svetlizky_brittle_2017}. In 
both situations, the stress profiles present the K-dominance predicted by LEFM 
for dynamic shear cracks with an associated dynamic energy release rate 
balancing the average fracture energy $G_c^{\rm ref}$. The details of the linear elastic 
stress solutions used in Fig.~\ref{fig:heterogeneous} are described in Appendix.

Few differences need to be commented; As $d^*$ significantly impacts the nucleation, dynamic rupture initiates under higher shear stress along interface \textit{A} than \textit{B} and consequently propagates at faster velocities. Both explain the different stress amplitudes between the two interfaces in Fig.~\ref{fig:heterogeneous}. The high frequency radiations visible in the stress profile of interface \textit{A} are another difference arising from the interplay of dynamic ruptures with heterogeneities larger than the process zone \cite{barras_interplay_2017}, and therefore mainly for interface \textit{A}. Nevertheless, their wavelength and amplitude are expected to decay for micro\-contacts smaller than the two orders of magnitude considered in our simulations and become out of the resolution of macroscopic experiments. Finally, additional differences could exist for 3-dimensional systems. Indeed, the in-plane distortions of the slip front caused by tough asperities larger than $d^*$ (as in configuration \textit{A}) could cause intense stress concentrations strongly impacting the overall rupture dynamics (as reported in the context of dynamic fracture \cite{dunham_supershear_2003,barras_supershear_2018-1}).

\section{Discussion and concluding remarks}

Between two realistic rough surfaces in contact, a dense spectrum of junction sizes forms the real contact area, which often
barely exceeds few percents of the apparent area of the contact plane \cite{dieterich_direct_1994}. The contacting asperities form clusters whose sizes typically follow a power-law distribution \cite{dieterich_imaging_1996}. Moreover, the strength of each asperity could vary following Gaussian or Weibull distribution.
In this context, our results predict the length under which the details of the microstructure can be homogenized along the tip of a nucleating slip patch. Interestingly, this length is equivalent to the characteristic junction size $d^*$ used to study the formation of wear particles \cite{aghababaei_critical_2016,frerot_mechanistic_2018}. Indeed, the 
strength of the junctions smaller than $d^*$ can be averaged (cf. responses of interface \textit{B} in 
Fig.~\ref{fig:onset:multicontacts}a), whereas the toughness of the microcontact junctions larger than $d^*$ are 
individually impacting the macroscopic frictional behavior of the interface (cf. responses of interface \textit{A} in 
Fig.~\ref{fig:onset:multicontacts}a). The combination of the criterion described in this paper with models simulating the contact of two rough surfaces \cite{yastrebov_infinitesimal_2015,frerot_fourier-accelerated_2019} open new prospects to investigate the frictional strength of contact interfaces. Such models could notably account for three-dimensional effects (e.g. shear-induced anisotropy \cite{sahli_shear-induced_2019}, shielding of neighboring rupture fronts \cite{aghababaei_asperity-level_2018} or its pinning by tough asperities \cite{gao_first-order_1989}).   

Any modification of the characteristic junction size $d^*$ (lubrication, coating) or the microcontact topography (sanding) will thereby impact the macroscopic frictional strength (even if such modifications are only visible at a microscale and do not change the average interface properties). The brittle-to-ductile transition discussed in this work brings then an interesting avenue to rationalize the ``slippery but tough'' behavior of lubricated interfaces discussed in the introduction. As reported by Bayart \textit{et al.} \cite{bayart_slippery_2016}, the lubrication significantly increases the critical slip distance $\delta_c$ and the interface fracture energy $G_c$. Moreover, a reduction of the interface adhesion also leads to an increase of the characteristic junction size $d^*$ \cite{aghababaei_effect_2019,brink_adhesive_2019}. Dry contact can hence be viewed as a strong but fragile interface, where slip initiates by a sharp concentration of the shear stress and damage zone at the edge of the microcontacts, followed by the abrupt brittle failure of individual microcontacts. After lubrication, the damage zone distributed over multiple microcontacts leads to the strength-dominated ductile failure of several junctions, resulting macroscopically into a more slippery yet tougher interface. 

Whereas the microcontacts topography together with $d^*$ play a significant role at nucleation, the macroscopic rupture dynamics appears to be much less impacted by the microscopic details and comply with the theoretical predictions for average homogenized properties. This observation is in good agreement with a recent set of frictional experiments revealing how the fracture energy inverted from interfacial displacements shows significant variations around the average and uniform value inverted from strain measurements in the bulk \cite{berman_dynamics_2020}.

More broadly, this work also find implications in our understanding of the failure of heterogeneous media, particularly in the context of multi-scale and hierarchical materials, for which the microstructure organization can be tuned to enhance the overall material properties 
\cite{munch_tough_2008,mirkhalaf_overcoming_2014}. 

\begin{acknowledgments}
This work was supported by the Swiss National Science Foundation (Grant 
No. 162569 ``Contact mechanics of rough surfaces'').
\end{acknowledgments}

\newpage

\section{APPENDIX}

\makeatletter 
\renewcommand{\thefigure}{A\@arabic\c@figure}
\renewcommand{\theequation}{A\@arabic\c@equation}
\makeatother
\setcounter{figure}{0}
\setcounter{equation}{0}    

\begin{figure}
\centering
\includegraphics[width=\linewidth]{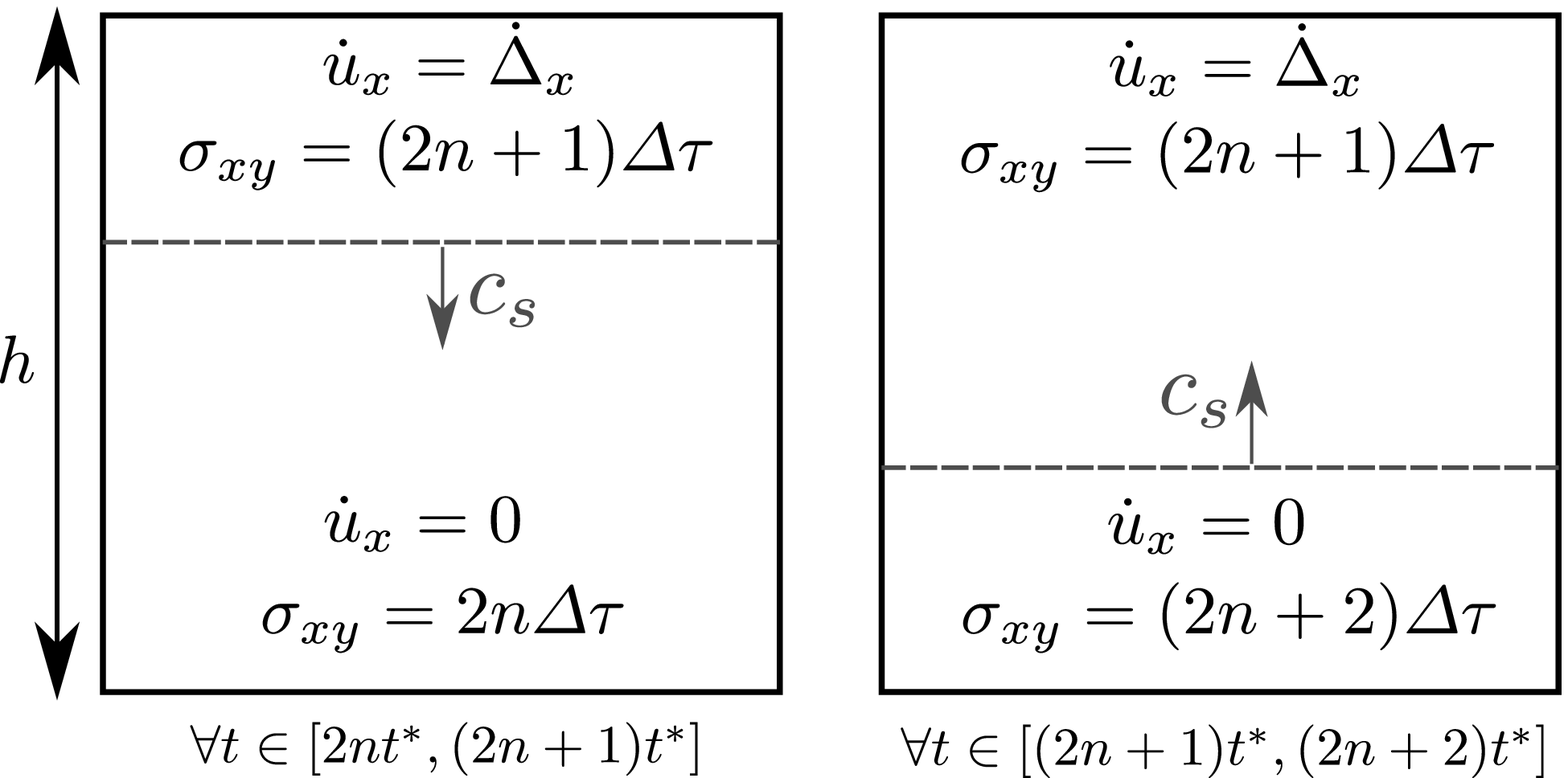}
\caption{\label{fig:step_loading} Elastodynamic solution in the absence of interfacial slip. The dynamic fields are mediated by the vertical propagation of a shear wave front characterized by $\varDelta\tau=\mu/c_s\;{\dot{\Delta}}_x$. $t^*=h/c_s$ is the time needed by the front to travel between the top and bottom surfaces and $n\in\mathbb{N}$ is the total number of reflections observed at the top boundary.}
\end{figure}

\appendix \section{End-member elastic solutions}
The numerical results presented in the manuscript are supported by theoretical solutions derived hereafter in the framework of linear elasticity which rests upon the following momentum balance equation:
\begin{equation}
\nabla \cdot \vec{\sigma}(x,y,t) = \rho\ddot{\vec{u}}(x,y,t).
\label{equ:mombalance}
\end{equation}
In the equation above, $\nabla$ is the divergence operator and we recall that $\vec{\sigma}$ is the Cauchy stress tensor, $\vec{u}$ the 
displacements vector and $\ddot{\square}$ denotes a double time derivative. At time $t=0$, the two continua presented in Fig.~\ref{fig:setup} are initially at rest and start being progressively loaded by a shear wave whose amplitude corresponds to $\varDelta\tau=\mu/c_s\;{\dot{\Delta}}_x$. $\mu$ is the elastic shear modulus and $c_s$ the shear wave speed such that $t^*=h/c_s$ is the wave travel time between the top and bottom surfaces. Figure~\ref{fig:step_loading} presents the elastodynamic solution of this system under the boundary conditions listed in Eq.~(\ref{equ:BC}). In this state of simple shear, the only non-zero components of $\vec{\sigma}$ are the shear stress $\sigma_{xy} = \sigma_{yx} = \mu \partial u_x/\partial y$ such that the elastic strain energy reduces to
\begin{equation}
 E_{\rm el} = \frac{1}{2\mu}\int_{-\frac{h}{2}}^{\frac{h}{2}}\int_0^l(\sigma_{xy})^2 dxdy.
 \label{equ:ele}
\end{equation}
Integrating the stress of the solution presented in Fig.~\ref{fig:step_loading} according to Eq.~(\ref{equ:ele}) leads to the quadratic build-up of strain energy depicted by the black dash lines in Figs.~\ref{fig:homogeneous}b and \ref{fig:homogeneous}c. 

After an initial loading phase, the build-up of strain energy is limited by the nucleation of slip and the progressive failure of the interface. As the system is initially at rest, the energy conservation implies that
\begin{equation}
 E_{\rm pot} + E_{\rm kin} + E_{\rm frac} = 0.
 \label{equ:ebalance}
\end{equation}
$E_{\rm pot} = E_{\rm el} - W_{\rm ext}$ is the potential energy, such that Eq.~(\ref{equ:ebalance}) can be rewritten after the complete interface failure as
\begin{equation}
 E_{\rm el} + E_{\rm kin} = W^*_{\rm ext} - E^*_{\rm frac}.
 \label{equ:ebalance2}
\end{equation}
As discussed in the manuscript, the right-hand-side terms of Eq.~(\ref{equ:ebalance2}) reaches constant values, respectively $W^*_{\rm ext}$ and $E^*_{\rm frac}$, while the left-hand-side terms represent an eventual excess of mechanical energy remaining in the system after the rupture. 

As function of the size of the region where sliding nucleates (i.e. the process zone size $l_{pz}$), two end-member situations exist. In the limit of an infinitesimally small process zone, this excess of mechanical energy can be related to the energy barrier governing the nucleation of a singular shear (mode II) crack. From Linear Elastic Fracture Mechanics (LEFM) \cite{griffith_phenomena_1921,irwin_analysis_1957,anderson_fracture_2005}, the rupture propagation starts according to the following thermodynamic criterion:
\begin{equation}
K_{II} > K_c.
\label{equ:lefm_crit}
\end{equation}
In the equation above, $K_c$ is the interface 
fracture toughness, which can be computed from the fracture energy as 
\begin{equation}
K_c = \sqrt{G_c\frac{E}{(1-\nu^2)}}.
\label{equ:kc}
\end{equation}
$K_{II}$ is the stress intensity factor, which depends on the far-field shear 
stress $\sigma^{\infty}_{xy}$, the initial crack size ($w_0$ in our setup) and a dimensionless factor $\chi$ accounting for the geometry:
\begin{equation}
K_{II} = \chi\sigma^\infty_{xy}\sqrt{\pi w_0}.
\label{equ:kII}
\end{equation}
In this manuscript, $\chi$ is approximated as $1.12$ for the edge crack configuration of interest \cite{anderson_fracture_2005}. The rupture is then expected to 
initiate when
\begin{equation}
\sigma_{xy} \geq \sigma^\infty_{xy}=\frac{1}{\chi}\sqrt{\frac{G_c}{\pi w_0}\frac{E}{(1-\nu^2)}}.
\label{equ:critical_stress}
\end{equation}
By assuming homogeneous shear stress within the two solids, the elastic strain energy required to initiate the rupture can be approximated by
\begin{equation}
E^{\rm lefm}_{\rm el} = \frac{1}{2\mu}\int_{-\frac{h}{2}}^{\frac{h}{2}}\int_0^l(\sigma^\infty_{xy})^2 dxdy = \frac{G_c}{\chi^2}\frac{hl}{\pi w_0(1-\nu)},
\label{equ:Elefm}
\end{equation}
which represents a strain energy barrier governing the onset of rupture growth. In the limit of a process zone larger than the length of the interface, the failure progressively occurs everywhere along the contact plane once the shear stress reaches the interface strength $\sigma_{xy} = \tau_c$ such that no energy barrier exists and $W^*_{\rm ext}=E^*_{\rm frac}$.

\begin{figure}
\centering
\includegraphics[width=\linewidth]{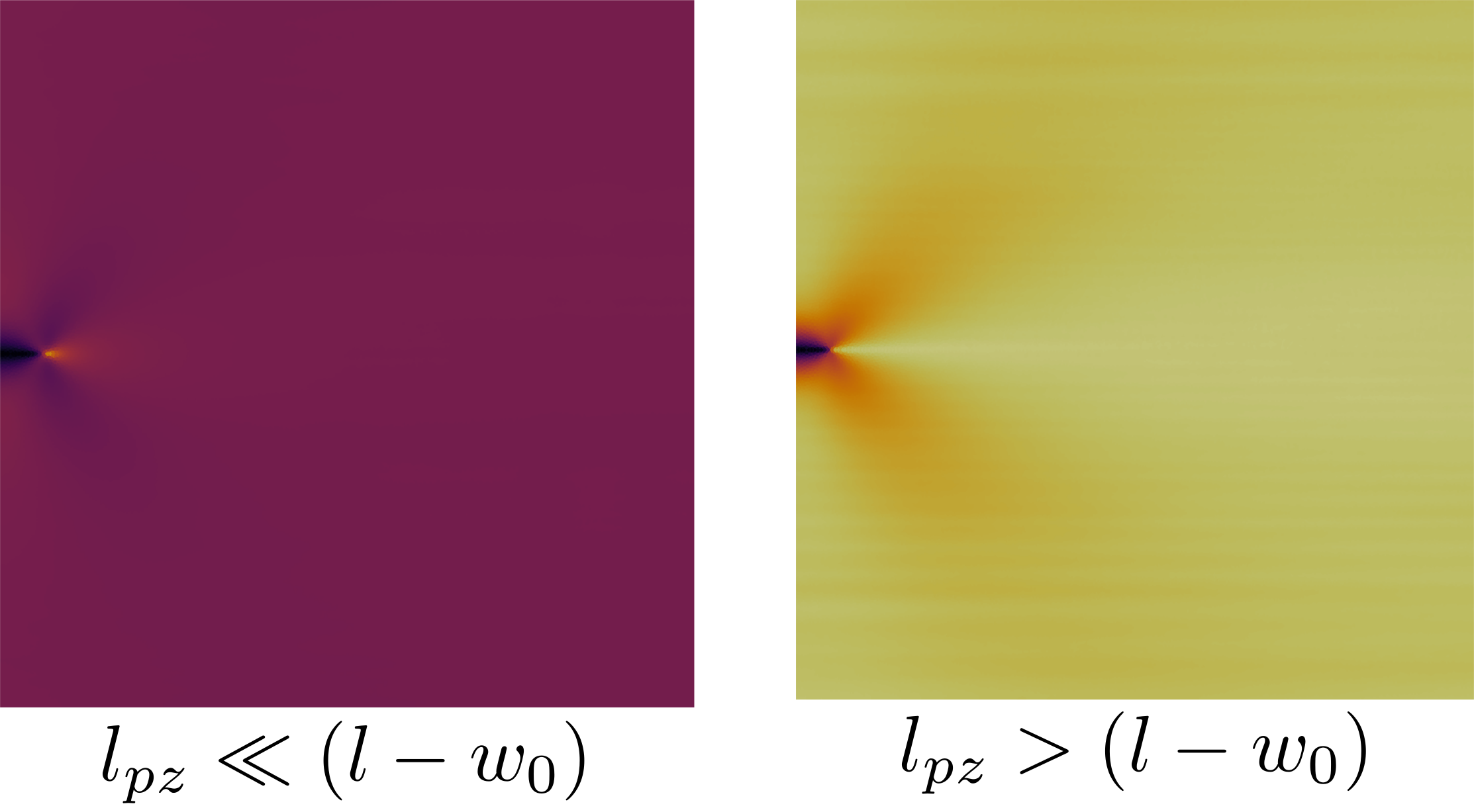}
\caption{\label{fig:lpz} Shear stress profiles before the onset of sliding for the two different failure mechanisms selected by the size of the process zone. In the left plot ($l_{pz}\ll (l-w_0)$), the stress concentrates at the very edge of the junction and the subsequent rupture corresponds to the sharp crack-like event studied in Fig.~\ref{fig:homogeneous}b. In the right plot ($l_{pz}>(l-w_0)$), the stress is uniform over the junction and leads to the ductile failure presented in Fig.~\ref{fig:homogeneous}c.}
\end{figure}

Figure \ref{fig:lpz} presents the shear stress profiles existing for these two end-member situations prior to the rupture. In the manuscript, this transition is studied in terms of the energy balance but the same approach could be used to predict the macroscopic force $F^*_{\textmd{ext}}$ required to trigger sliding, i.e. to reach the interface frictional strength. Invoking that the dynamic effects are negligible before the onset of sliding, two end-member solutions can be similarly derived for $F^*_{\textmd{ext}}$. In the limit of an infinitesimal process zone ($l_{pz}\ll (l-w_0)$), the force is controlled by the far-field shear stress predicted by LEFM and corresponds to

\begin{equation}
F^{\rm lefm}_{\textmd{ext}} = \sigma^\infty_{xy}\cdot l=\frac{l}{\chi}\sqrt{\frac{G_c}{\pi w_0}\frac{E}{(1-\nu^2)}}.
\label{equ:F_lefm}
\end{equation}
Conversely, if $l_{pz}>(l-w_0)$ the applied force should balance the peak strength along the entire contact junction such that $F^*_{\textmd{ext}}$ approaches

\begin{equation}
F^{\rm str}_{\textmd{ext}} = \tau_c\cdot(l-w_0).
\label{equ:F_strength}
\end{equation}

\begin{figure}
\centering
\includegraphics[width=\linewidth]{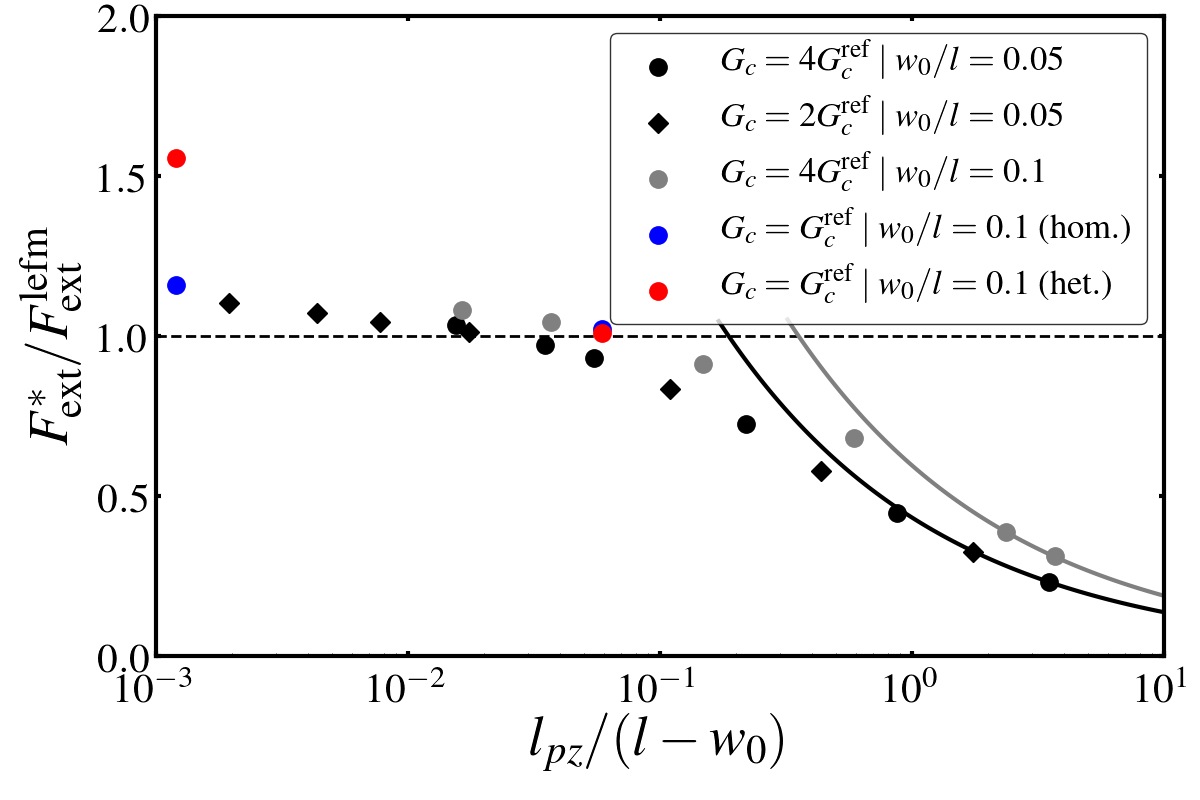}
\caption{\label{fig:force} External force required to trigger sliding as function of the process zone size for the simulations reported in the Figs.~\ref{fig:homogeneous}a and \ref{fig:onset:multicontacts}a of the manuscript. In the large process zone limit, the data follows $F^{\rm str}_{\textmd{ext}}$, whose evolution predicted by Eq.~(\ref{equ:F_str2}) is depicted by the black and grey solid lines for the two studied geometries, respectively $w_0/l=0.05$ and $w_0/l=0.1$. With shorter process zone sizes, the external force saturates at $F^{\rm lefm}_{\textmd{ext}}$, the value predicted from brittle fracture. The blue and red dots present the values observed for respectively the homogeneous and heterogeneous large-scale simulations. Please refer to the presentation in the main text for more information about the different setups.}
\end{figure}

\appendix \section{Numerical method}

\begin{figure*}[t!]
\centering
\includegraphics[width=\linewidth]{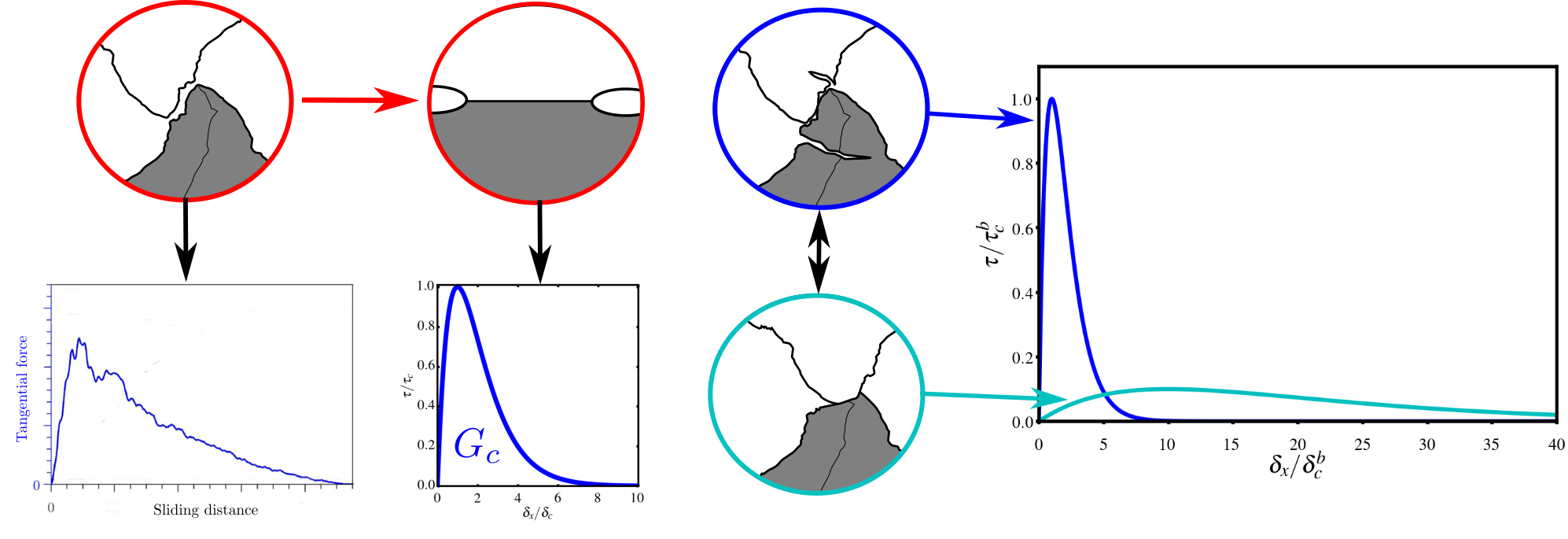}
\caption{\label{fig:coarse_graining} From left to right: Typical force versus 
slip profile observed during the shearing of two interacting asperities in 
molecular dynamics simulations (see \cite{aghababaei_debris-level_2017} for a detailed 
presentation of the method and setup). Such behavior can be conveniently described by the exponential cohesive law given in 
Eq.~(\ref{equ:exponential_law}) and derived from a Rose-Ferrante-Smith 
\cite{rose_universal_1981} type of universal binding potential. The exponential cohesive law allows for saving the cost of describing 
the fine details of asperity contact and, in return, the coarse-grained junctions can embed the microcontacts failure behavior into the macroscopic response of frictional systems. More notably, these coarse-grained junctions also reproduce the essential observations of the molecular dynamics simulations: the brittle-to-ductile transition in the failure of microcontact junctions controlled by an identical 
characteristic length scale (see the discussion in Section \textbf{Characteristic length scales of the brittle-to-ductile failure transition} of the manuscript). Examples of a very brittle cohesive law in dark blue ($\tau^b_c$; $\delta^b_c$) and a more ductile one in cyan ($\tau^d_c=0.1\tau^b_c$; $\delta^d_c=10\delta^b_c$) having the same fracture energy.}
\end{figure*}

The elastodynamic equation (Eq.~(\ref{equ:mombalance})) is solved with a finite element approach using 
a lumped mass matrix coupled to an explicit time integration scheme 
based on a Newmark-$\beta$ method \cite{newmark_method_1959}. The stable time 
step is defined as function of the dilatational wave speed $c_d$ and the 
spatial discretization $\Delta s$ as 
\begin{equation}
\Delta t=0.7\frac{\Delta s}{c_d},
\end{equation}
with $\Delta s$ being typically set to $\frac{l}{1000}$ in this work. For the large simulations of interfaces with a 
heterogeneous microstructure, the discretization is brought to $\frac{l}{5000}$ leading to about 70M degrees of freedom. 
The virtual work contribution of the frictional plane is written as
\begin{equation}
 \hat{W}(t) = \int_0^l\tau(x,t)\hat{\delta}_x(x,t)dx,
\end{equation}
with $\hat{\square}$ denoting a ``virtual'' quantity and $\delta_x(x,t)=u_x(x,0^+,t)-u_x(x,0^-,t)$ being the interfacial slip between the top and bottom surfaces. The shear traction acting at the interface $\tau$ is assumed to derive from an 
exponential Rose-Ferrante-Smith universal potential $\Phi$ \cite{rose_universal_1981} and is expressed as
\begin{equation}
 \tau = \frac{\partial\Phi}{\partial\delta_x} 
= \frac{\delta_x}{\delta_c}\tau_ce^{1-\frac{ 
\delta_x} {\delta_c}}.
\label{equ:exponential_law}
\end{equation}
In Eq.~(\ref{equ:exponential_law}), $\tau_c$ and $\delta_c$ are respectively the maximum strength and critical 
slip of the interface characterizing the exponential \textit{traction-separation} law sketched in 
Fig.~\ref{fig:coarse_graining}, for which the fracture energy corresponds to
\begin{equation}
G_c = \int_0^{\infty}\tau d\delta_x = e\tau_c\delta_c.
\label{equ:fracture_energy}
\end{equation}
Modeling the failure of the junctions existing between two rough surfaces 
motivates the choice of the exponential potential and associated cohesive law 
(Eq.~(\ref{equ:exponential_law})). Indeed, Aghababaei \textit{et al.} 
\cite{aghababaei_critical_2016,aghababaei_debris-level_2017,
aghababaei_asperity-level_2018} used atomistic simulations to study the shear 
failure of various kinds of interlocking surface asperities and reported how the 
evolution of the profile of the ``far-field`` tangential force versus sliding 
distance follows a similar evolution than the exponential cohesive law (see for 
example Fig.~1 of \cite{aghababaei_debris-level_2017}). In this context, the 
chosen cohesive formulation should be understood as a generic ''coarse-grained`` description of the failure of the underlying microcontact 
junctions. This idea is illustrated in Fig.~\ref{fig:coarse_graining}.
Interestingly, this coarse-grained formulation is, at the same 
time, representative of the micromechanical behavior of microcontact 
junctions and similar to the slip-weakening description of friction used in 
the macroscopic modeling of contact planes 
\cite{andrews_rupture_1976-1,svetlizky_properties_2016,barras_interplay_2017}.
The main objective of this work is to study the nucleation process, but the model could add residual friction at the valleys or in the trail of the fronts with no loss of generality.

Capturing the multi-scale nature of the problem requires an efficient and scalable parallel implementation of the finite element method, capable of 
handling several millions of degrees of freedom on high-performance computing clusters. To this aim, we use our homemade open-source finite element software \textit{Akantu}, whose implementation is detailed in \cite{richart_implementation_2015,vocialta_3d_2017} and whose sources can be 
freely accessed from the \textit{c4science} platform \footnote{{h}ttps://c4science.ch/project/view/34/}.
More details about the finite element formulation \cite{belytschko_nonlinear_2014,hughes_finite_2000,zienkiewicz_finite_2005} and 
the implementation of cohesive element models \cite{xu_void_1993,ortiz_finitedeformation_1999} can be found in the reference 
papers. 

\appendix \subsection{Material properties}

The results are discussed in the manuscript with adimensional scales but the 
material properties of Homalite used in the simulations are given to the reader 
for the sake of reproducibility: Young's modulus $E = 5.3$ [GPa], Poisson's 
ratio $\nu = 0.35$, shear wave speed $c_s=1263$ [m/s], and reference interface 
fracture energy $G_c^{\rm ref} = 23$ [J/m$^2$].

\subsection{Dynamic fracture mechanics}
For a detailed presentation of the dynamic fracture theory, the reader is 
redirected to the reference textbooks 
\cite{freund_dynamic_1990,kostrov_principles_1988,ravi-chandar_dynamic_2004}. 
For a mode II shear crack moving at speed $v_c$, the dynamic energy balance is 
expressed from the dynamic stress intensity factor $K_{II}$ and a universal 
function of the crack speed $A_{II}$:
\begin{equation}
 G_c = G = \frac{1-\nu^2}{E}K_{II}^2A_{II}(v_c),
\end{equation}
with
\begin{equation}
 A_{II}(v_c) = \frac{\alpha_s v_c^2}{(1-\nu)Dc_s^2},
 \label{equ:a2}
\end{equation}
where $\alpha_{s,d}^2=1-v_c^2/c_{s,d}^2$, and $D=4 \alpha_d \alpha_s - 
(1+\alpha_s^2)^2$.
As for the static crack depicted in Fig.~\ref{fig:setup}, stresses 
immediately ahead of a dynamic front are dominated by a square-root 
singular contribution. The latter can be expressed in a polar system of 
coordinates $(r,\theta)$ attached to the crack tip and as function of the 
dynamic stress intensity factor $K_{II}$ \cite{freund_dynamic_1990}:
\begin{align}
\sigma_{xx} &= 
-\frac{K_{II}}{\sqrt{2\pi 
r}}\frac{2\alpha_s}{D}\Big\lbrace 
(1+2\alpha^2_d-\alpha^2_s)\frac{\sin\frac{1}{2}\theta_d}{\sqrt{
\gamma_d}}\nonumber\\
&-(1+\alpha_s^2)\frac{\sin\frac{1}{2}\theta_s}{\sqrt{\gamma_s}}
\Big\rbrace, \nonumber\\
\sigma_{xy} &= 
\frac{K_{II}}{\sqrt{2\pi 
r}}\frac{1}{D}\Big\lbrace 
4\alpha_d\alpha_s\frac{\cos\frac{1}{2}\theta_d}{\sqrt{
\gamma_d}}-(1+\alpha_s^2)^2\frac{\cos\frac{1}{2}\theta_s}{\sqrt{\gamma_s}}
\Big\rbrace, \label{equ:modeII_stress_sing}\\
\sigma_{yy} &= 
\frac{K_{II}}{\sqrt{2\pi 
r}}\frac{2\alpha_s(1+\alpha^2_s)}{D}\Big\lbrace 
\frac{\sin\frac{1}{2}\theta_d}{\sqrt{\gamma_d}}-\frac{\sin\frac{
1}{2}\theta_s}{\sqrt{\gamma_s}}\Big\rbrace, \nonumber
\end{align}
with $\gamma_{s,d}=\sqrt{1-(v_c\sin\theta/c_{s,d})^2}$ and 
$\tan\theta_{s,d}=\alpha_{s,d}\tan\theta$.
\newline
\newpage
\noindent 
The good agreement with LEFM predictions reported in Fig.~\ref{fig:heterogeneous} is obtained with 
\begin{equation}
K_{II}=\sqrt{\frac{G_c^{\rm ref} E}{(1-\nu^2)A_{II}(v_c)}}
\end{equation}
and by seeking for the position of the front $x_{\rm{tip}}$ and its 
propagation velocity $v_c$ that give the best predictions of the simulated stress profiles according to a nonlinear least-squares regression 
\cite{jones_scipy_2001,watson_levenberg-marquardt_1978,branch_subspace_1999}. Just as in Williams series describing static cracks \cite{williams_stress_1957}, non-singular contributions could be added to describe stresses evolution far from the tip (cf. region 
(III) in Fig.~\ref{fig:setup} of the main manuscript) following the approach presented in \cite{svetlizky_classical_2014}. The non-singular contribution has however a limited influence on the resulting mapping shown in Fig.~\ref{fig:heterogeneous}.

\bibliographystyle{apsrev4-2}
%\bibliography{bibliography.bib}

\end{document}